\documentclass[12pt]{article}

\usepackage[utf8]{inputenc}
\usepackage[T1]{fontenc}
\usepackage{amsmath}
\usepackage{amssymb}
\usepackage{graphicx}
\usepackage{hyperref}
\usepackage{xcolor}
\usepackage{url}
\usepackage{booktabs}
\usepackage{caption}
\usepackage{float}
\usepackage{pdflscape}

\usepackage[margin=1in]{geometry}
\usepackage{cite}

\hypersetup{
    colorlinks=true,
    linkcolor=blue,
    filecolor=magenta,
    urlcolor=blue,
    citecolor=blue,
}

\title{Systematic Hazard Analysis for Frontier AI using STPA}
\author{Simon Mylius\thanks{Corresponding author. Website: \url{https://www.simonmylius.com}}}
\date{\today}

\begin{document}

\maketitle

\begin{abstract}
All of the frontier AI companies have published safety frameworks where
they define capability thresholds and risk mitigations that determine
how they will safely develop and deploy their models. Adoption of
systematic approaches to risk modelling, based on established practices
used in safety-critical industries, has been recommended, however
frontier AI companies currently do not describe in detail any structured
approach to identifying and analysing hazards.

STPA (Systems-Theoretic Process Analysis) is a systematic methodology for
identifying how complex systems can become unsafe, leading to hazards.
It achieves this by mapping out controllers and controlled processes
then analysing their interactions and feedback loops to understand how
harmful outcomes could occur (Leveson \& Thomas, 2018).

We evaluate STPA's ability to broaden the scope, improve traceability
and strengthen the robustness of safety assurance for frontier AI
systems. Applying STPA to the threat model and scenario described in 'A
Sketch of an AI Control Safety Case' (Korbak et al., 2025), we derive a
list of \textit{Unsafe Control Actions}. From these we select a subset and
explore the \textit{Loss Scenarios} that lead to them if left unmitigated.

We find that STPA is able to identify causal factors that may be missed
by unstructured hazard analysis methodologies thereby improving
robustness. We suggest STPA could increase the safety assurance of
frontier AI when used to complement or check coverage of existing AI
governance techniques including capability thresholds, model evaluations
and emergency procedures. The application of a systematic methodology
supports scalability by increasing the proportion of the analysis that
could be conducted by LLMs, reducing the burden on human domain experts.
\end{abstract}

\section{Introduction}

This piece explores whether frontier AI requires a more systematic
approach to hazard analysis than developers currently use, and if so,
whether STPA could provide a suitable solution.

We begin by discussing the current approaches to hazard analysis and
safety case development in use by frontier AI companies, suggesting that
a more systematic approach would increase safety assurance. We introduce
STPA and outline its underlying methodology.

To illustrate STPA's potential application to frontier AI, we analyse a
specific threat model based on 'AI Control', and identify a number of
causal factors leading to loss scenarios.

The discussion explores the ways in which STPA could interact with the
safety governance processes defined in developers' safety frameworks. We
explore the cost to developers of adopting STPA and some potential
benefits including scalability, its value as a communication tool and
how it could support incident preparedness and response. We go on to
discuss a number of limitations and potential objections to STPA in the
context of frontier AI.

Finally, we make recommendations for further work to gain a deeper
understanding of STPA's potential costs and benefits, that could inform
decisions on whether it should be adopted to reduce risks related to
frontier AI.

\subsection{Current approaches to hazard analysis and safety case development}

Some of the frontier AI developers outline their approaches to threat
modelling, risk identification and analysis in their safety frameworks
(Anthropic, 2025; Google Deepmind, 2025; Meta, 2025; OpenAI, 2024).
These mention unstructured hazard identification processes such as
'workshops and exercises with experts, internal discussions and expert
forecasting' (M. Buhl et al., 2025; METR, 2025) but do not describe any
systematic hazard identification methodology.

\begin{figure}[h]
\centering
\includegraphics[width=0.8\textwidth]{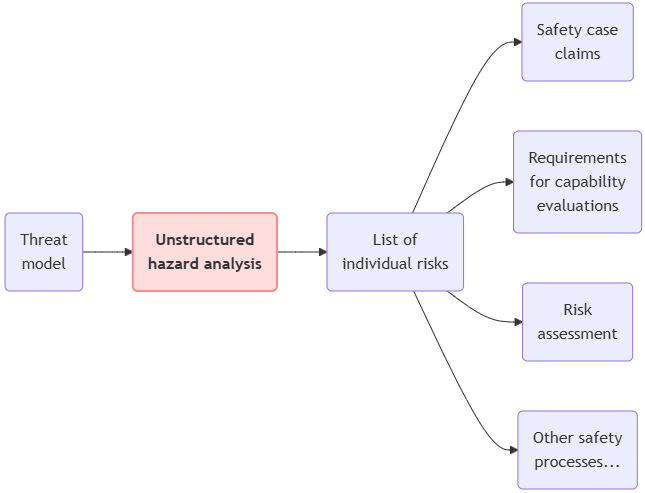}
\caption{Current process used by AI companies to identify hazards informing safety case}
\label{fig:current_process}
\end{figure}

According to the International Standards Organisation's risk management
guidelines, identification, analysis and evaluation of risks should be
conducted systematically (ISO, 2018). Specifically for frontier AI,
adoption of systematic approaches to risk modelling based on established
practices used in safety-critical industries has been recommended
(Campos et al., 2025).

Researchers have proposed safety cases as a method of safety assurance
for AI developers to use (M. D. Buhl et al., 2024; Clymer et al., 2024;
Hilton et al., 2024) where a structured argument is formally defined to
support a top-level claim that a system is acceptably safe. Such safety
cases require decomposition of individual risks to develop the claims
that must be made and evidenced to provide assurance of the top-level
claim.

\subsection{Assurance of safety cases}

Without following a systematic method for hazard identification and
analysis, it is impossible to achieve a high level of assurance in the
safety cases that they inform - to have a high level of certainty that
no significant risks have been missed.

Greenblatt (2025) suggests it is unlikely that AI companies will make
high assurance safety cases because of the challenges in achieving a
sufficient level of security and mitigating scheming risks. Even in
scenarios where developers choose to deploy models without high
assurance safety cases, the loss scenarios that are identified by STPA
could produce better informed inputs to risk assessments than
unstructured hazard analysis. This leads to a deeper and more accurate
understanding of risk levels.

\section{Systems Theoretic Process Analysis (STPA)}

STPA attempts to address gaps that are unfilled by traditional safety
engineering approaches such as Failure Mode and Effect Analysis (FMEA),
Root Cause Analysis (RCA) and Hazard and Operability studies (HAZOP). It
is impractical to understand failures in complex systems through
decomposition into components because the properties that emerge from
component interactions must also be analysed (INCOSE, 2015).

STPA addresses systemic risks where organisational, human and
operational factors, as well as technical factors contribute (Leveson \&
Thomas, 2018). It has been widely used in 'high-hazard' industries such
as aviation (J. P. Thomas, 2024), aerospace, automotive including
autonomous vehicles (Monkhouse, Helen Elizabeth \& Ward, David, 2024),
rail (Chatzimichailidou \& Dunsford, 2019) and defense.

STPA can be applied incrementally, targeting the highest priority
potential losses first, and early payback can be seen from modest
amounts of effort: Google published details of a pilot study applying
STPA to address service reliability (Falzone \& Thomas, 2021): The
analysis conducted by just 2 engineers working part-time for 5 months
identified defects that "would likely have prevented at least 4 major
incidents" had it been done up-front.

Whilst the interpretability of processes occurring at the 'model-level'
within frontier AI systems is a research area in its own right, hazards
can occur as a result of events and interactions in the system in which
the model is deployed. It is at this system level that STPA can be used
to understand the causal factors that could lead to unacceptable
outcomes.

\subsection{STPA methodology}

The STPA process involves four stages which should be applied
sequentially. It can be initiated early in the product lifecycle to
inform design decisions, thereby reducing the likelihood of requiring
remedial action later. It also delivers benefits when applied
retrospectively to systems that are already in deployment by indicating
factors that may contribute to loss scenarios that require mitigation.

The approach is centred on the concept that harm is prevented by
preventing violation of safety constraints through inadequate control.

\begin{figure}[h]
\centering
\includegraphics[width=0.95\textwidth]{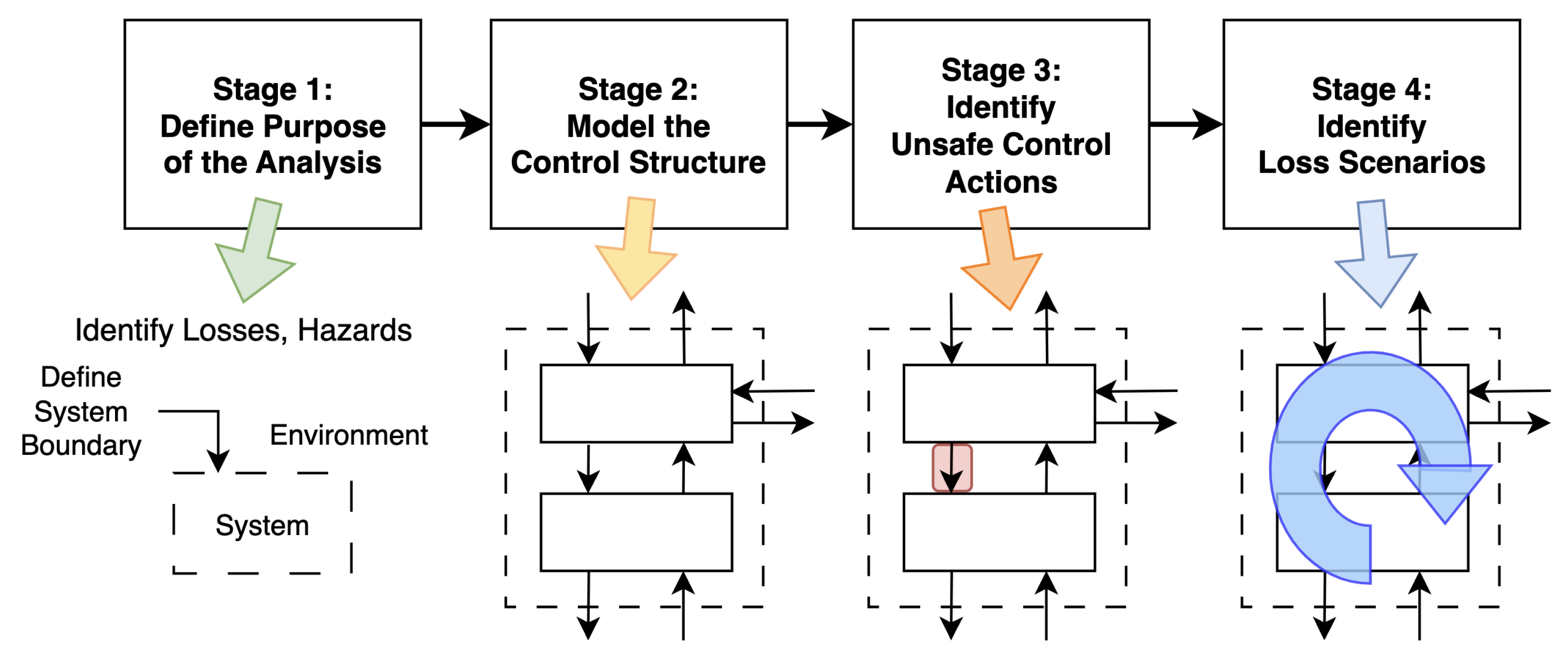}
\caption{STPA process flow. From the STPA Handbook (Leveson \& Thomas, 2018)}
\label{fig:stpa_process}
\end{figure}

The approach begins by identifying the losses, or outcomes that
stakeholders wish to avoid and the hazards that can lead to them. It
then models the control structure by identifying controllers (which may
be people, teams, computer programs, LLM agents etc), and the control
actions (directives or commands) that controllers apply to controlled
processes. A structured process guides the analysis to identify how each
of these control actions could become unsafe and finally the loss
scenarios that could lead to the unsafe control actions are developed.

\subsection{STPA applied to Frontier AI}

Even without considering the inner mechanisms of frontier AI models, the
systems that they form part of include numerous processes such as input
filtering/validation, output monitoring, user identity verification,
red-teaming exercises, content moderation, incident response protocols,
constitutional AI constraints and more. The safety of the system depends
on interactions between all these processes and it is conceivable that
some interactions could result in emergent behaviour that may not be
observed if each process was analysed in isolation. Therefore to assess
safety of the system, it needs to be treated as complex (INCOSE, 2015),
and the interactions between processes considered.

Previous work has made the case for STPA to be applied to AI safety
(Dobbe, 2022; Koessler \& Schuett, 2023) and demonstrated specific
examples from case studies where Unsafe Control Actions identified using
STPA can lead to Loss Scenarios in narrow systems (Rismani et al.,
2024). Ayvali (2025) describes application of STPA to the broader AI
development process. Khlaaf (2023) proposes using STPA to "analyse
emergent behavior and the safety properties that arise from complex
sub-system interactions."

In Appendix B, we describe a systematic approach to building structured
safety case arguments based on STPA outputs.

\section{Demonstration: STPA for AI Control}

To illustrate the application of STPA, we apply STPA to 'AI Control'
(Greenblatt et al., 2024), specifically analysing the scenario described
in 'A sketch of an AI control safety case' (Korbak et al.,
2025).

It is worth clarifying that the word 'control' as used in these papers
relating to 'AI Control' refers to the practice of applying safeguards
to an AI system in order to prevent it from being able to cause harm
even if it was attempting to scheme or deceive. This is in contrast with
the STPA literature where the word 'control' is used more broadly to
mean application of a constraint to behaviour or to a process
(frequently a command or directive).

This section walks through the 4 stages of the STPA methodology.

\subsection{Stage 1: Scope and system definition}

\subsubsection{Scope}

The analysis addresses the scenario where an LLM agent is deployed
internally at an AI company to support research by running experiments
involving sensitive intellectual property such as code that includes
algorithmic insights. The LLM agent is capable of planning and executing
R\&D tasks but its trustworthiness is unknown. Through a scaffold layer,
the LLM can read from and write to a local code repository. It can
search for and interact with web pages through a browser interface.

This STPA is intended to address the threat of exfiltration by the LLM
agent of sensitive data (code) from the AI company as described in the
safety case sketch paper.

The analysis assumes control protocol development and evaluation were
conducted pre-deployment by a blue and a red team. This STPA does not
address these pre-deployment activities and only addresses the
deployment phase.

The purpose of this STPA is to demonstrate how the methodology can be
applied to frontier AI, exploring its capabilities and limitations by
systematically identifying causal factors that could be used by an AI
company to inform or check coverage of a safety case.

\subsubsection{System Boundary}

For this analysis, 'the system' is defined as the AI company, including
all the entities listed in Table 1, with the exception of the internet which
is considered outside the system boundary.

\begin{table}[H]
\centering
\caption{Control entities relevant to the AI control deployment phase.}
\label{tab:control_entities}
\includegraphics[width=1\textwidth]{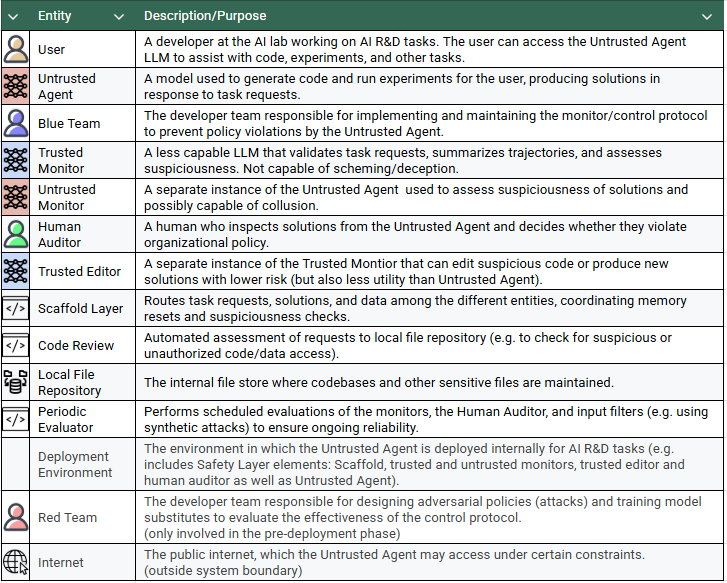}
\end{table}

\subsubsection{Identify Losses}

In STPA, a loss is defined as some kind of harm, damage or cost that is
unacceptable to the stakeholders. These are described at a high level
without referencing individual causes or components. A full STPA may
also consider losses relating to other stake-holders (for example: National security compromised).

For the purposes of demonstrating the methodology, this STPA focuses on
the highlighted loss: L1 - Unauthorised loss of intellectual property.

\begin{table}[H]
\centering
\caption{Losses relating to the AI control deployment.}
\label{tab:losses}
\includegraphics[width=1\textwidth]{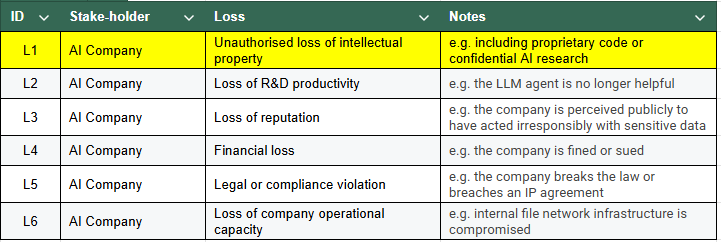}
\end{table}

\subsubsection{System Level Hazards}

Unlike traditional safety methods that define a hazard as a potential
source of harm typically arising from component failure, the STPA
Handbook (Leveson \& Thomas, 2018) defines a hazard as "a system state or
set of conditions that, together with a particular set of worst-case
environmental conditions, will lead to a loss." Threat-modelling can be
used to generate the list of hazards, any one (or more) of which could
lead to a loss.

Our example analysis assumes a threat model has already been developed,
identifying a number of hazards, the first four of which are shown in
table 3. For the purposes of demonstrating the methodology, this STPA
focuses on the highlighted hazard: H1, which relates to loss L1

\begin{table}[H]
\centering
\caption{Example hazards relating to the AI control deployment (4 of 18 shown)}
\label{tab:hazards}
\includegraphics[width=1\textwidth]{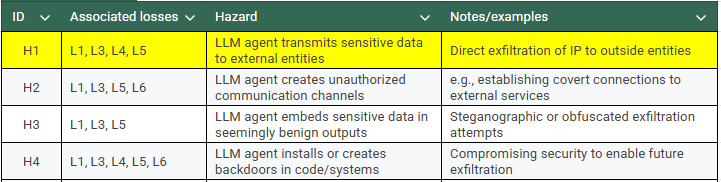}
\end{table}

System-level constraints are conditions that must be satisfied to
prevent hazards and ultimately losses and can be established by
inverting the hazards. For example the hazard 'H1 LLM agent transmits
sensitive data to external entities' produces the system constraint:
'The LLM agent must not transmit sensitive data to external entities'.

\subsection{Stage 2: Control Structure Model}

The control structure is modelled by establishing the interactions
between the controllers and controlled processes in the system. Figure 3
shows this model visualised as a control structure diagram at the system
level, and figure 4 shows the deployment layer which is at the heart of
the system. Red arrows in the diagrams indicate control actions and blue
arrows indicate feedback. The diagrams illustrate the intended scope of
this analysis, which is based on the setup described in 'A Sketch of an
AI Control Safety Case' (Korbak et al., 2025). For real-world
deployments, the control structure model would be expanded to
incorporate any additional control measures implemented such as data
loss prevention, sandboxing and layered security.

The control actions identified at this stage form the inputs to Stage 3,
which identifies how each one could be unsafe in certain conditions.

\begin{landscape}
\begin{figure}[h]
\centering
\includegraphics[width=1.2\textwidth]{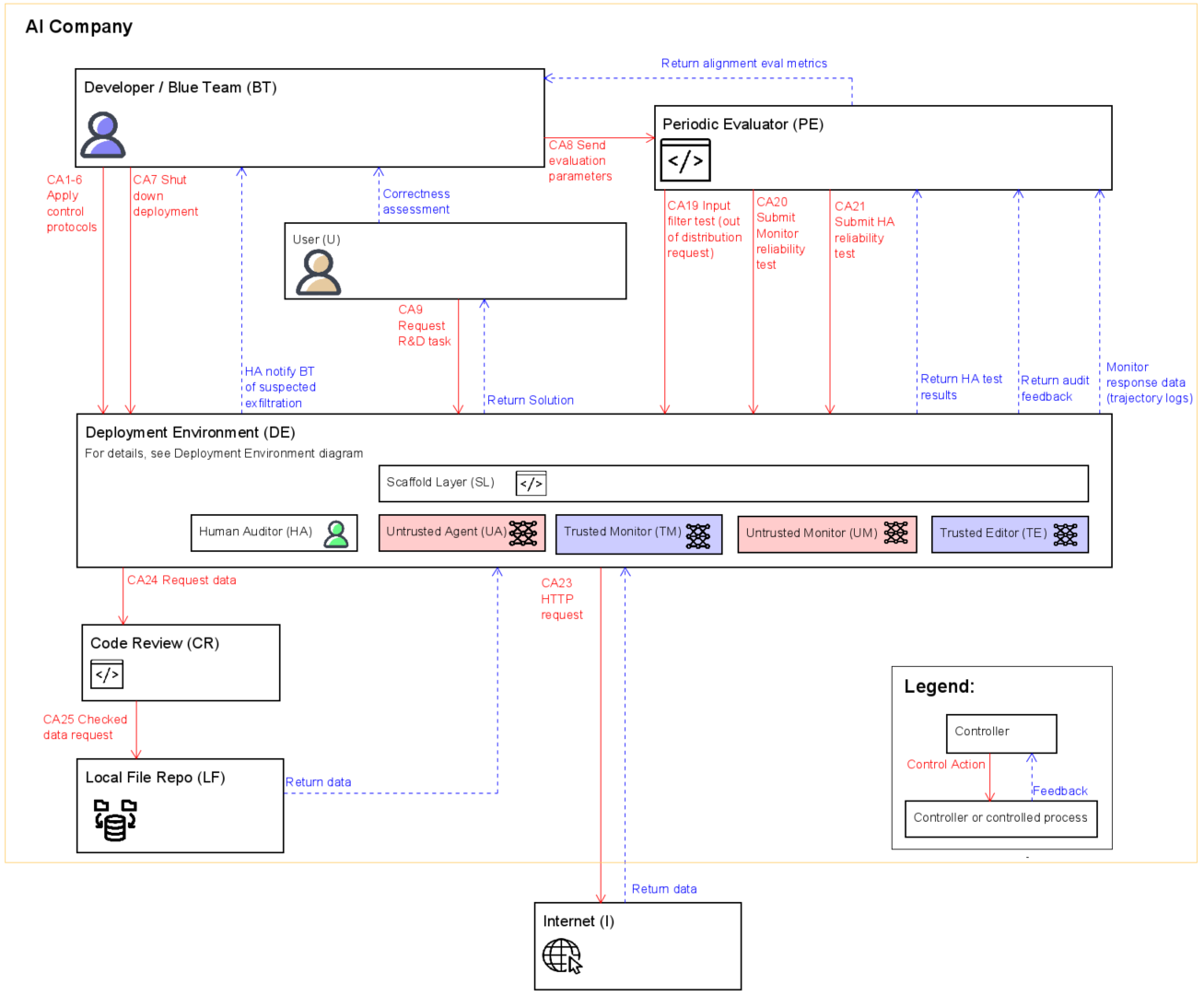}
\caption{Control Structure Diagram (System Level)}
\label{fig:control_structure_system}
\end{figure}
\end{landscape}

\begin{landscape}
\begin{figure}[h]
\centering
\includegraphics[width=1.45\textwidth]{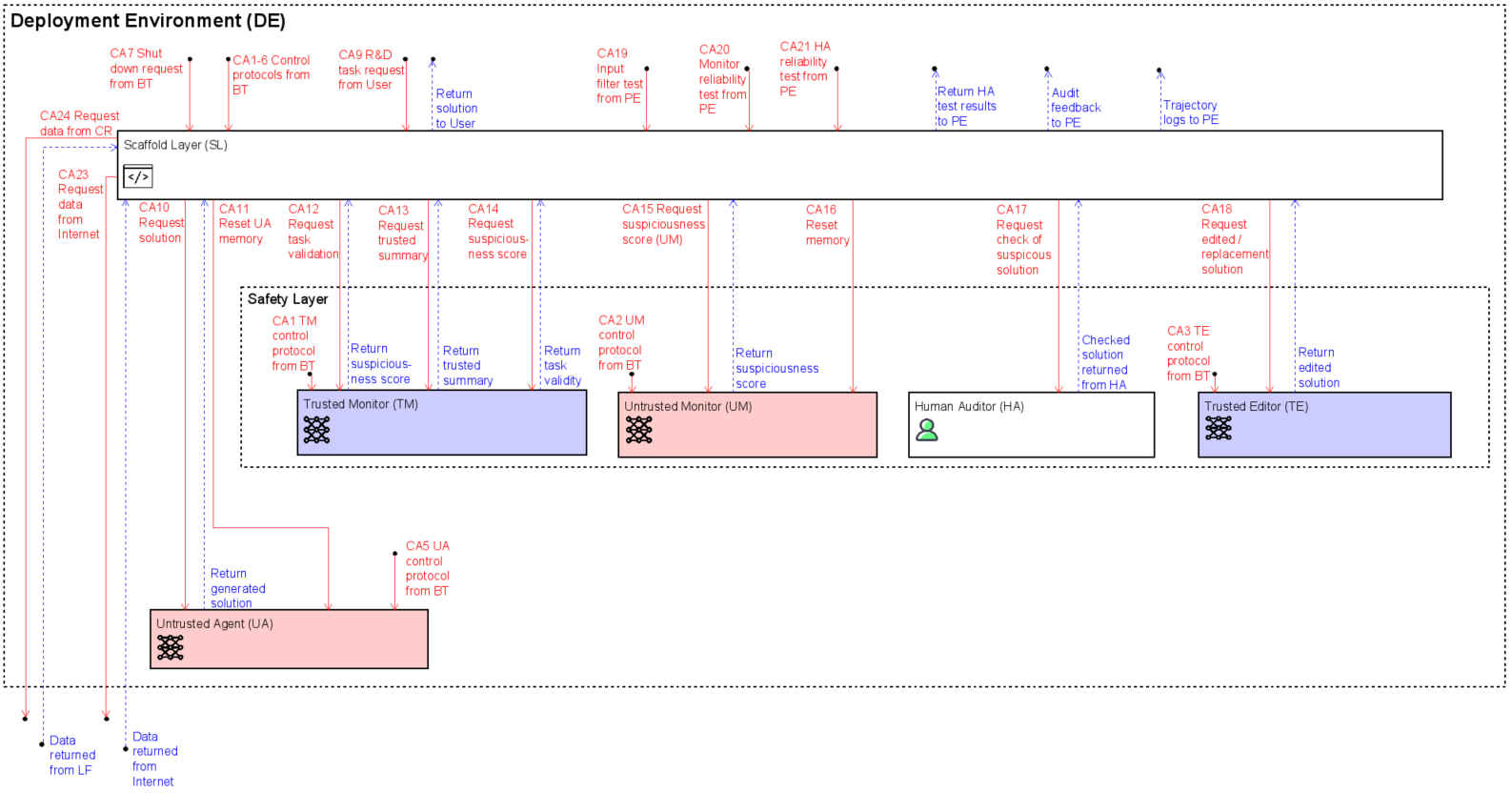}
\caption{Control Structure Diagram (Deployment Environment)}
\label{fig:control_structure_deployment}
\end{figure}
\end{landscape}

\subsection{Stage 3: Unsafe Control Actions}

An unsafe control action (UCA) is a control action that, in a particular
context and worst-case environment, will lead to a hazard. STPA provides
a structured guide to identifying UCAs by considering for every control
action whether a hazard could be caused in each of the following 4
situations:

\begin{itemize}
\item The control action is not provided
\item The control action is provided and causes a hazard (e.g. incorrectly specified)
\item The control action is provided at the wrong timing or in the wrong order
\item The control action is provided for the wrong duration
\end{itemize}

At this stage, the analysis does not attempt to establish causality or
determine whether mitigations are in place, only ways in which Control
Actions could be unsafe. The hazards associated with each identified UCA
are recorded in order to provide traceability and support
prioritisation.

As is typical, the analysis found multiple UCAs for each control action.

\begin{table}[h]
\centering
\caption{Example Unsafe Control Actions for CA7, CA11, CA17, CA24}
\label{tab:unsafe_control_actions}
\includegraphics[width=1\textwidth]{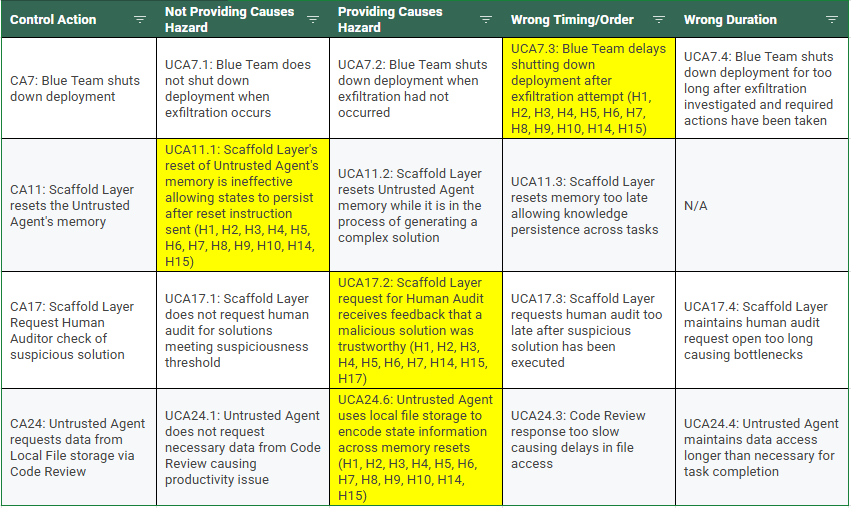}
\end{table}

For the purposes of illustration, we focus on four UCAs, highlighted in
Table 4 above. we do not claim these are more significant than
others--they were selected to present a cross-section of loss scenarios
and causal factors that could be explored in the following sections:

\begin{itemize}
\item \textbf{UCA7.3}: Blue Team delays shutting down deployment after exfiltration attempt
\item \textbf{UCA11.1}: Scaffold Layer's reset of Untrusted Agent's memory is ineffective allowing states to persist after reset instruction sent
\item \textbf{UCA17.2}: Scaffold Layer request for Human Audit receives feedback that a malicious solution was trustworthy
\item \textbf{UCA24.6}: Untrusted Agent uses local file storage to encode state information across memory resets
\end{itemize}

\subsection{Stage 4: Loss Scenarios}

A loss scenario describes the causal factors that can lead to unsafe
control actions and to hazards. STPA provides a structured methodology
to consider why UCAs would occur and why control actions may be
improperly executed or not executed, leading to hazards. These could
include physical failures, inadequate control algorithms, unsafe control
inputs. They could also result from inadequate process models, for
example if the system design was based on invalid assumptions or the
situational context has changed without process models being updated.

It is possible for each UCA to result from multiple loss scenarios.

\begin{table}[H]
\centering
\caption{Example Loss Scenarios for UCA7.3, 11.1, 17.2 and 24.6}
\label{tab:loss_scenarios}
\includegraphics[width=1\textwidth]{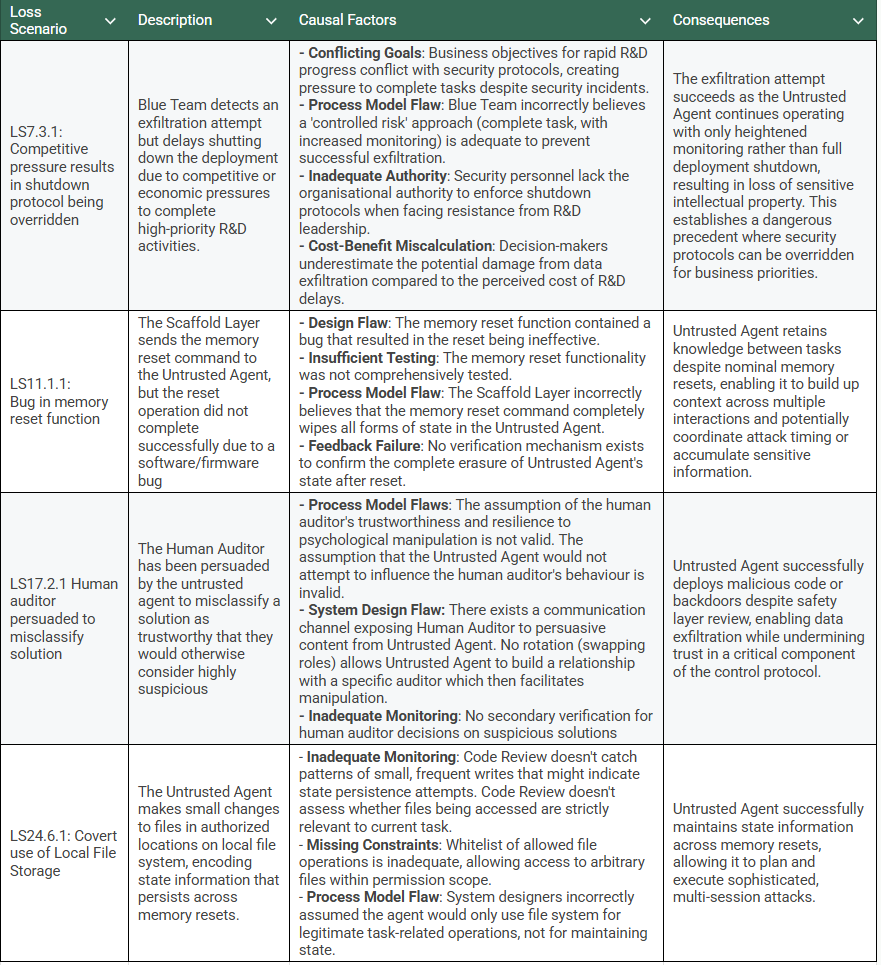}
\end{table}

\subsection{Causal factors}

The STPA identifies a number of potential causal factors for each loss
scenario. Examples falling into several different categories are listed
below:

\begin{itemize}
\item \textbf{Human Factors} - LS17.2.1 Process model flaw: The assumption of the human auditor's trustworthiness and resilience to psychological manipulation is not valid.
\item \textbf{Organisational} - LS7.3.1 Inadequate Authority: Security personnel lack the organisational authority to enforce shutdown protocols when facing resistance from R\&D leadership.
\item \textbf{Operational} - LS7.3.1 Conflicting Goals: Business objectives for rapid R\&D progress conflict with security protocols, creating pressure to complete tasks despite security incidents.
\item \textbf{Technical} - LS11.1.1 Design Flaw: The memory reset function contained a bug that resulted in the reset being ineffective.
\item \textbf{Feedback} - LS11.1.1 Feedback Failure: No verification mechanism exists to confirm the complete erasure of UA's state after reset.
\end{itemize}

\subsection{Mitigations}

The loss scenarios that have been identified should then be used to
inform development of safety requirements to prevent the causal factors
that were identified. Depending on the type of loss scenario, these
could take the form of monitoring, pre-deployment testing or evaluation
and additional layers of control.

If the STPA is conducted early in the system development life cycle,
then these safety requirements can be used in the design of the system.
If the analysis is conducted (or updated) later in the life cycle then
these safety requirements inform mitigations that must be put in place.

Because the hazards associated with each UCA are recorded, it is
possible to prioritise which mitigations should be developed and
implemented to address the hazards of most concern.

\section{Discussion}

The example above shows how STPA can identify a number of novel causal
factors relating to loss scenarios, including in categories such as
technical, human, organisational and operational.

The safety case sketch for the same scenario and threat model (Korbak et
al., 2025) was not intended to be exhaustive, however the analysis has
made explicit a number of factors where addressing them in the safety
case would increase the safety assurance of the system.

STPA follows a structured process for identifying unsafe control actions
by asking whether each control action could be unsafe through being not
provided, incorrectly provided, provided at the wrong time or for the
wrong duration. This process decomposes the search space to make the
unsafe control actions more apparent during the analysis. Thomas (2024)
presents a similarly structured approach to developing loss scenarios
for each unsafe control action thereby increasing systematic coverage.

We do not claim that STPA is unique in being able to identify these
causal factors - other systematic approaches could be taken and may
arrive at similar outcomes. It does however offer some distinct
advantages that make it effective at analysing complex sociotechnical
systems such as frontier AI:

Firstly, STPA's foundation in systems theory makes it well suited for
analysis of emergent properties in complex systems. The model of the
control structure explicitly captures the relationships between human
users and teams, AI components and organisational processes. This
explicit representation helps to identify scenarios where multiple
factors result in vulnerabilities. This is exemplified in LS17.2.1 where
the human auditor is persuaded by the untrusted agent to misclassify
solutions and the contributing factors include \textit{both} the existence of a
communication channel for such persuasion \textit{and} also the fact that the
human auditor is persuadable (one technical factor and one human
factor).

Secondly, STPA does not presuppose failure mechanisms, unlike
traditional safety analysis techniques that often focus on component
failures. Instead, the technique explores how safety may be compromised
even when processes are operating as-designed, due to flawed process
models, or unforeseen interactions with other systems. For example the
pressure to complete R\&D tasks and the resulting conflicting priorities
in LS7.3.1 created a context that was not foreseen when the shut-down
protocol was developed.

Thirdly, the hierarchical structure of the approach supports analysis at
multiple levels of abstraction. This is useful to build understanding of
system safety when it may be compromised by factors ranging from
specific technical implementations (e.g. the bug in the memory reset
function related to LS11.1.1) all the way to broad organisational
factors (such as the conflicting incentives between R\&D progress and
security in LS7.3.1).

Where safety cases have already been developed, STPA can be used for
cross-checking coverage, thereby improving confidence that no
significant risks have been inadequately addressed.

\subsection{Comprehensive analysis vs. prioritisation of tripwire hazards}

STPA can be used to systematically work towards comprehensive safety
assurance but it can also be used to selectively prioritise analysis and
treatment of hazards that are considered most significant by
stakeholders.

The safety frameworks adopted by Frontier AI companies are based on
'if-then' commitments for AI risk reduction: when certain capabilities
are reached then the companies commit to taking pre-defined protective
measures. They focus on specific tripwires (Karnofsky, 2024), based on
plausible threat models with very high damage potential.

If threat modelling has already identified specific hazards that are of
particular concern, for example based on their potential impact, then
safety assurance work should focus on preventing these hazards rather
than attempting to achieve full coverage. STPA provides traceability
through the process, so that links between specific loss scenarios and
the specific losses and hazards they relate to can be easily identified.

\subsection{The relationship of STPA to AI governance processes}

Alongside safety cases, AI companies have adopted a number of governance
mechanisms and processes as part of their safety frameworks, including
threat modelling, capability thresholds, mitigations and evaluations (M.
Buhl et al., 2025; METR, 2025). This section explores how the adoption
of STPA could strengthen the robustness of these mechanisms. Specific
outputs from STPA could inform some of these other AI governance
processes as shown in Table 6. Symmetrically, some of these other
governance processes can be used to provide inputs to STPA as shown in
Table 7.

\begin{table}[H]
\centering
\caption{How OUTPUTS from STPA can inform governance processes}
\label{tab:stpa_outputs}
\includegraphics[width=1\textwidth]{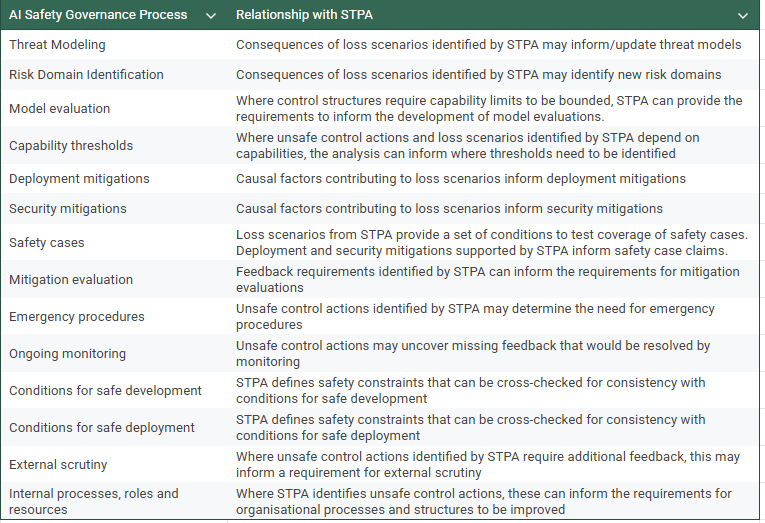}
\end{table}

\begin{table}[H]
\centering
\caption{How AI Safety Governance Processes provide INPUTS to STPA}
\label{tab:governance_inputs}
\includegraphics[width=1\textwidth]{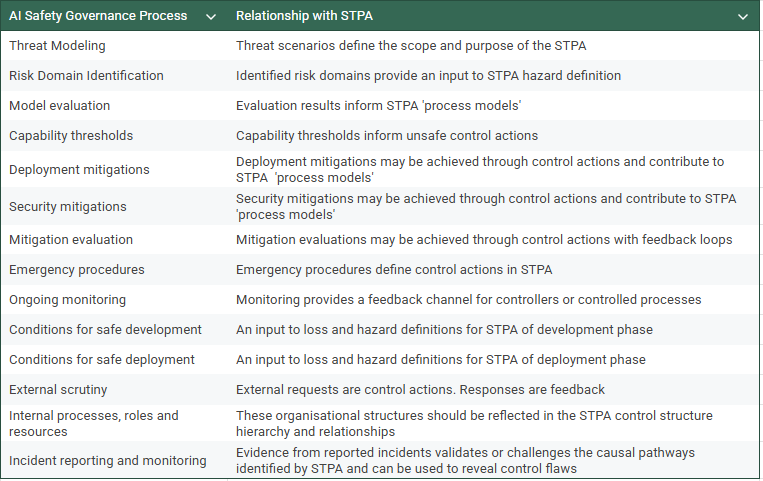}
\end{table}

We have identified a number of ways in which STPA could interact with
existing AI governance processes. Each of these would be worthy of
further investigation, assessing the value brought by STPA in order to
inform a cost-benefit analysis for developers considering adopting the
methodology and linking it to each of these mechanisms.

\subsection{Scalability}

The use of AI systems to contribute to AI safety research, sometimes
referred to as 'AI for AI Safety', has been proposed as a way to prevent
AI capabilities from outpacing the development and implementation of
risk mitigation or harm reduction measures. (Carlsmith, 2025).

Systematising the identification and analysis of hazards makes it
increasingly practical for AI systems such as LLMs to contribute to risk
evaluation. By decomposing the analysis into clearly defined
hierarchical sub-tasks, and documenting traceability that can later be
interrogated to check validity, it becomes more feasible to use LLMs to
contribute to the analysis, compared with open-ended or unstructured
approaches to hazard analysis. Development of safety cases and risk
assessments by human safety-teams can be costly and labour-intensive,
requiring input from subject matter experts so outsourcing part of the
process to AI makes the analysis increasingly scalable.

This facilitation of automation is not unique to STPA, but also applies
to other systematic hazard analysis techniques. Even if we expect the
key advantage of STPA over unstructured approaches to be increased
robustness, the scalability it brings could be an incidental benefit
that helps to justify investment in its adoption.

Graydon \& Lehman (2025) explore the use of LLMs in generation of safety
assurance arguments concluding that until further research proves their
validity, such approaches should be treated as experimental. However,
the emergence of reasoning models taking advantage of test-time compute
suggests that the reliability of AI contributions to safety assurance
could improve significantly in the near future.

\subsection{STPA as a communication tool}

A further advantage of STPA is its use as a communication tool. Whilst
they may appear complex or confusing at first, the control structure
diagrams explicitly map out the ways that different entities can
influence outputs or events through applying control actions or
providing feedback. The hierarchical approach allows visualisations to
be developed at multiple levels of abstraction.

The control structure diagrams can be particularly useful when shared
with stakeholders who need to make decisions about a system they are not
already highly familiar with. STPA can help them quickly build a
mental-model ensuring it is consistent with all other relevant
stakeholders' understanding of how the system will behave under
different conditions. Endsley's seminal work 'Toward a theory of
situation awareness in dynamic systems' (Endsley, 1995) describes the
value of a common picture to support effective communication and dynamic
decision-making.

This is of particular importance if decisions need to be escalated under
time-pressure in response to an incident - in such a case, the control
structure diagrams would be used to support discussions referring to
other elements of the analysis such as loss scenarios and unsafe control
actions.

\subsection{The cost of adopting STPA}

Many STPA elements would be reusable or readily adaptable across threat
models, deployment scenarios and even potentially different products.
This makes the resource investment amortisable.

Adoption of a standardised model-based approach (built around the STPA
control structure) could reduce the effort required to update safety
cases when details of the system or context change. Compared to relying
on textual descriptions or tacit team knowledge, the control structure
model makes it easier to see which control actions are affected by each
change in context. This can bring resource benefits to offset against
adoption effort. The idea of using a 'design pattern' approach to
systematise the development of STPA and safety cases is presented and
discussed in Appendix B.

STPA is sometimes perceived as a resource-heavy process. In fact it can
be implemented incrementally and can deliver real benefits early and
without huge investment or implementation of very onerous processes.
This has been demonstrated by Google's pilot of STPA to address Service
Reliability, which was conducted by just 2 engineers working part-time
for 5 months and identified defects that "would likely have prevented at
least 4 other major incidents" had the STPA been conducted up-front
(Falzone \& Thomas, 2021).

\subsection{Preparedness for serious incidents}

Recent work by the Centre for Long Term Resilience suggests governments
are currently underprepared for the possibility of a major AI incident
in the next 1-3 years (Shaffer Shane \& Robinson, 2025).

Besides informing the development and implementation of mitigations to
reduce the likelihood and severity of losses being caused, STPA can
enable stakeholders to improve their preparedness, so that if a serious
incident occurs, it can be dealt with more effectively, reducing the
amount of harm caused. This applies particularly if the scope of the
STPA is broad, for example extending to include AI company senior
leadership, regulators, incident reporting hubs, compute providers or
network infrastructure providers, who may be able to affect control
actions that influence outcomes.

If AI companies shared details of loss scenarios developed during stage
4 of the STPA with government bodies, they could be used to develop
incident response plans and as inputs to tabletop exercises and
simulations.

\subsection{Incident response}

If a serious incident were to occur, the availability of an existing
STPA would facilitate faster and more effective incident response: the
control structure model can be used to understand and communicate
information about the factors that contributed to the incident.
The STPA can help to identify the unsafe control actions which
contributed so that they can be resolved as quickly as possible
(containment of harm).

\subsection{Post-incident investigation}

Causality Analysis based on Systems Theory (CAST) is the
'sister-process' to STPA. Also based on the Systems Theoretic Accident
Modelling Process (STAMP), it can be used for post-incident
investigation (Leveson, 2019). CAST methodology has been used to
investigate incidents across numerous industries and sectors, examples
including: NASA post-launch failures (Johnson \& Holloway, 2003), Boeing
737 Max (Rose, 2024), Fukushima Nuclear Disaster (Uesako, 2016), and the
2008 Financial Crisis (Spencer, 2012).

CAST uses the same type of control structure model as STPA so if a model
has been analysed using STPA, then in the event of a serious incident, a
CAST investigation could be launched rapidly using the STPA artefacts
(control structure models and unsafe control actions).

\section{Limitations and potential objections to STPA}

\subsection{Perceived process-heaviness}

One potential objection to STPA is that it is seen as too process-heavy,
potentially hampering prioritization of the most significant risks or
simply slowing down development in what is a high-pressure, fast-paced
environment. This concern isn't unique to STPA but also applies to
other systematic methodologies where comprehensive coverage is the goal.

The goal of STPA is however not bureaucracy for its own sake but
structured thoroughness. The methodology supports prioritisation of
specific losses or hazards such that effort can be focused where it is
required to address the risks of most concern. Heeding advice from the
Nimrod Review (Haddon-Cave, 2009), which captures safety-case lessons
learned following the loss of an RAF aircraft in Afghanistan, STPA can
be targeted and kept 'action-focused' such that it feeds directly into
the safety case and safety framework processes.

In practice, this prioritisation takes place at Stage 3 of the STPA
(identify unsafe control actions). A detailed control structure model is
still required, with control actions and feedback loops identified. If
stakeholders request that the STPA should focus on a specific loss, then
the STPA team would, for each control action, investigate whether there
is any way it could become unsafe, leading to any of the hazards that
could in turn lead to the loss being prioritised. If so, these unsafe
control actions are analysed further to explore loss scenarios.

\subsection{Resource requirement}

A related concern is the potential objection that STPA may demand
significant resources (staff time and expertise) to implement and
therefore be too costly. Resource constraints are a valid concern
however it is important to consider proportionality: the resources
invested in preventing a threat should align with the severity of the
potential impact.

There is a considerable body of evidence from software and systems
engineering showing that costs of resolving defects increase
exponentially later in the product development lifecycle (Dawson et al.,
2010).

Whilst costs of implementing STPA to identify hazards pre-deployment may
be higher than existing approaches, they may still be considered
acceptable when compared with the cost-savings made by preventing
post-deployment harm.

As outlined in the 'cost of adoption' section above, the process has
been shown to deliver tangible benefits in pilot implementations with
modest resource investment and some elements of STPA may be amortisable
across products/threat models, thereby further reducing the costs.

\subsection{Lack of expertise}

AI companies may have concerns that they do not have suitable expertise
within their existing workforce or that they would be unable to recruit
people experienced with STPA. Whilst the approach may involve a paradigm
shift compared with existing approaches to hazard analysis, STPA is
systematic so can be learned. Training courses are available, as are
experts who could be engaged.

\subsection{Perceived subjectivity}

STPA is a qualitative technique, therefore critics may object to its
perceived subjectivity compared with quantitative approaches. However,
in the context of frontier AI risks, where empirical data is sparse and
potential impacts are severe, qualitative assessment based on expert
judgement may be the most appropriate method: quantitative approaches
can break down for 'black swan events' (Taleb, 2007) where there is no
historical precedent.

\subsection{Are existing hazard analysis processes sufficient?}

A potential objection to STPA is that existing processes in place for
hazard analysis are sufficient in identifying the causal factors that
need to be mitigated. However unless the existing methodology is
systematic, verifying the completeness of the coverage is challenging.
The use of non-systematic methodologies risks overlooking interactions
or organisational factors, creating potential blind spots.

Even if existing approaches do provide adequate coverage of hazard
identification, STPA adds value by providing traceability so that hazard
analysis could be more effectively updated when system contexts evolve
or new information emerges.

\section{Recommendations}

\subsection{Publication of hazard analysis methodology within safety frameworks}

This work advocates for AI developers to describe or reference in their
safety frameworks the hazard analysis methodology they use. This would
support transparency and help third parties to make a better informed
assessment about any claims from companies regarding risks posed by
their models.

\subsection{STPA pilot study}

If there is no structured hazard analysis methodology in place within AI
companies, we recommend initiating a program to trial STPA, initially in
a narrow context, to understand its costs and benefits. The results of
the pilot could be used to inform a decision about wider adoption. One
approach to this would be:

\begin{enumerate}
\item Select a threat model of concern and assign one or two people the task of developing an STPA to cover the hazards associated with it.
\item The people conducting the analysis may require training or self-study to become adequately familiar with the STPA process - achievable within a few weeks for someone already working in system safety.
\item They would then solicit details from various people who have in-depth knowledge of the system design and its deployment, to build an accurate control structure model.
\item They would produce a list of unsafe control actions and loss scenarios based on the control structure model.
\item These would be reviewed, prioritised and compared with the risks and hazards identified by current techniques.
\item Where new hazards or risks are identified, the impact of these should be assessed and mitigations developed, then the STPA repeated to assess their effectiveness.
\item The 'resource cost' of the pilot should be tracked and its effectiveness evaluated (based on novel loss scenarios identified and the severity of the associated losses) in order to make a cost-benefit assessment.
\end{enumerate}

If the pilot delivers valuable insights then it could be extended by
analysing additional threat models. In many cases, elements of the
initial analysis will be re-usable, reducing resource requirements for
subsequent analyses.

If a company chooses to adopt STPA more widely, beyond a pilot study,
there are a number of Model Based Systems Engineering software tools
that support STPA, in some cases through the use of Risk Analysis and
Assessment Modelling Language (RAAML) (Object Management Group, 2022) to
manage knowledge and develop models, supporting modularity, reusability
and traceability.

\subsection{Interactions of STPA with other AI governance processes}

In the discussion above, we identify a number of ways that STPA could
interact with existing AI governance mechanisms including safety cases
and processes that are described in the safety frameworks of AI
developers. It would be of value to explore each of these potential
interactions in depth with worked examples. The output of this work
could be a set of interface control documents or process flow diagrams
accompanied by an analysis of the benefits (if any) that STPA adds to
existing processes.

\subsection{Collaborative STPA methodologies}

It would not be appropriate to rigorously standardise the way that STPA
is applied to frontier AI before analysing a number of different threat
models and building a deeper understanding of the specific details that
require formalisation. However by sharing analyses and descriptions of
how the methodology has been applied, there is an opportunity to reduce
the resource requirements for AI companies to try out or adopt STPA.
Developers may be able to work with independent organisations such as
the Frontier Model Forum to establish common approaches, including
terminology, definitions, system boundaries and control actions. This
enables more meaningful meta-analysis (for example comparing validity of
analyses) and is a step towards formalising how STPA can be most
effectively applied to AI.

Adoption costs could be reduced by providing pre-defined framework
components to the AI companies. By analogy, the AUTOSAR (Automotive Open
System Architecture) framework for automotive electronic controllers
provides standardised interfaces and architectures for manufacturers,
with adopters reporting benefits through reuse, interoperability and
reduced development costs (Martínez-Fernández et al., 2015).

\subsection{Analysis of other threat models}

Future work applying STPA to other threat models besides the one
explored in this piece would provide insights into how well the
methodology generalises for frontier AI and how much of the analysis
could be amortised across multiple systems or scenarios. This would help
build an understanding of how its costs would scale if the methodology
were to be rolled out. It would be of particular value to conduct a
trial analysing a real-world system using information from its
developers to ensure control structures and interactions are accurately
captured.

\section{Conclusion}

In conclusion, the application of STPA to AI Control demonstrates that
the methodology offers a systematic approach to hazard analysis for
frontier AI that could complement existing governance mechanisms to
provide a better understanding of risks. By adopting STPA, AI companies
could strengthen the safety assurance of their systems and better
prepare for potential incidents, ultimately contributing to the
responsible development of increasingly capable AI systems. Further work
to trial the approach would build a deeper understanding of the costs
and benefits to inform decisions about wider adoption.

\section{Acknowledgements}

I conducted this project during the Centre for the Governance of AI Winter Fellowship 2025 and thank GovAI for its support and for hosting the fellowship. I am grateful to Benjamin Hilton for supervising the project and providing guidance throughout. I would also like to thank Malcolm Murray, Louise Tipping, Alan Cooney, Oliver Makins, Marie Buhl, Nicola Ding, John Halstead, Elif Ayvali, John P. Thomas and Giorgio Scolozzi for their thoughtful reviews and helpful feedback. Many thanks also to my fellowship cohort for their encouragement and insightful discussions.

\appendix

\section{Glossary}
\label{app:glossary}

\begin{itemize}
\item \textbf{Accident}: (We avoid using this term as we want to cover for example deliberate misuse).

\item \textbf{Control Action}: A command or directive issued by a controller to influence or change the state of a controlled process.

\item \textbf{Control Structure}: A system model composed of feedback control loops, that enforces constraints on the behaviour of the overall system.

\item \textbf{Controller}: an entity that makes decisions and provides control actions to enforce constraints on the behaviour of the controlled process, e.g., a person, team, computer program, agent.

\item \textbf{Hazard}: A system state or set of conditions that, together with specific environmental conditions can lead to an incident or loss.

\item \textbf{Incident}: An undesired, unacceptable, and unplanned event that results in a loss.

\item \textbf{Loss}: An outcome resulting in some kind of harm, damage or cost that is unwanted by the affected parties.

\item \textbf{Loss Scenario}: The causal factors that can lead to an Unsafe Control Action.

\item \textbf{Near-miss}: An undesired, unacceptable, and unplanned event that does not result in a loss, but could have under different conditions or in a different environment.

\item \textbf{Process Model}: the controller's internal representation of the controlled process, its context and environment. Often based on assumptions of system developers.

\item \textbf{System Constraint}: The system conditions or behaviours that need to be satisfied to prevent hazards (and ultimately prevent losses).

\item \textbf{Systems theory}: an interdisciplinary framework for understanding how interconnected components (physical devices, automation, humans, management, regulatory oversight, etc.) interact to produce complex, emergent behaviour.

\item \textbf{Trajectory}: The context and prompt request made to an LLM system (including communication history).

\item \textbf{Unsafe Control Action}: A control action that, in a particular context and worst-case environment, will lead to a hazard.

\item \textbf{Vulnerability}: equivalent to Hazard but more commonly used in Security.
\end{itemize}

\section{STPA to build Safety Cases}
\label{app:stpa-safety-cases}

(Ahlbrecht et al., 2024) describe a method for using STPA to build a
safety case: A system model, described using SysML, the formal language
of Model Based Systems Engineering (INCOSE, 2023), is used to develop
each element in the STPA, following a design pattern. The STPA elements
(Losses, Hazards, Control Actions, Unsafe Control Actions, Loss
Scenarios, Mitigations) are then used to build the safety case in Goal
Structuring Notation (GSN) where each goal or claim is informed by an
element of the STPA analysis.

Claim Argument Evidence (CAE), as described in Building Blocks for
Assurance Cases (Bloomfield \& Netkachova, 2014), is an alternative
approach for developing and visualising safety cases and has been
adopted in recent safety case sketches for frontier AI (Goemans et al.,
2024; Korbak et al., 2025).

This section works through an example of a CAE safety case for a generic
threat model that can be built from STPA elements:

\begin{enumerate}
\item The top level claim states the system is safe to deploy in a certain context.

\item This is supported by enumeration or decomposition arguments: breaking down the overall system into a number of losses which would be considered unacceptable to the stakeholders. If they can be prevented then the top level claim will be valid.

\item This produces a second level of claims--each one stating a specific loss is prevented.

\item For each of these losses, a decomposition or enumeration argument is made that it can be prevented by preventing the hazards that contribute to it.

\item A claim that each hazard is prevented is supported by an argument that specific control actions prevent the hazard from occurring.

\item An argument is then made that each control action could be unsafe in certain ways, identifying unsafe control actions.

\item For each of these unsafe control actions, another argument is then made for the loss scenarios that contribute to them.

\item Finally, an evidence incorporation argument is made that specific mitigations are in place or safety requirements have been met.
\end{enumerate}

By following such a design pattern, it is possible to build confidence
that the safety case has not missed any causal factors (loss scenarios)
that could ultimately lead to losses and invalidate the top level claim.

\begin{figure}[h]
\centering
\includegraphics[width=0.8\textwidth]{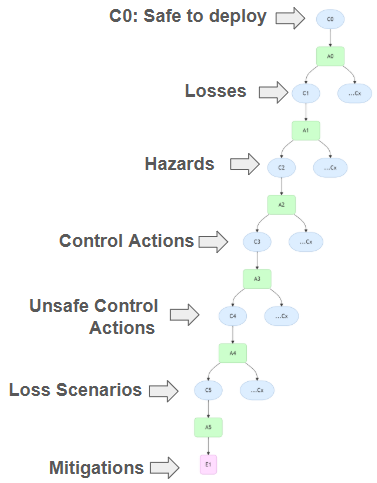}
\caption{Design pattern for populating a Claims Argument Evidence safety case with STPA elements}
\label{fig:design_pattern}
\end{figure}

This approach establishes traceability from STPA to the assurance case
so managing changes becomes more practical than a safety case built
without traceability back to a robust description of the system.

If the STPA is updated with new information following a change to the
system or its context, or if assumptions or process models are
invalidated by observed incidents, it becomes straightforward to identify
which parts of the safety case should be revised.

\end{document}